\documentclass{aa}  
\newcommand\dscript[2]{\bgroup#1\egroup_\bgroup#2\egroup}
\usepackage{graphicx}

\usepackage{amsmath,amssymb} 
\usepackage[flushleft]{threeparttable}

\newcommand*{\astrosun}{{\odot}}

\newcommand*{\earth}{{\oplus}}

\usepackage{txfonts}
\usepackage{xcolor}
\usepackage{ulem}

\newcommand\sredit[1]{{{#1}}}
\begin{document}

  \title{X-ray variability of the triplet star system LTT1445 and evaporation history of the exoplanets around its A component}
    
    \author{S. Rukdee
          \inst{1}
          \and
          J. Buchner\inst{1}
          \and 
          V. Burwitz\inst{1}
          \and
          K. Poppenh{\"a}ger\inst{2}
          \and
          {B. Stelzer}\inst{3}
          \and
          P. Predehl\inst{1} 
          }

   \institute{Max Planck Institute for Extraterrestrial     Physics, Giessenbachstrasse 1, 85748 Garching, Germany
         \and
         Leibniz-Institute for Astrophysics Potsdam (AIP), An der Sternwarte 16, 14482 Potsdam, Germany
         \and
         Institut f{\"u}r Astronomie und Astrophysik, Eberhard Karls Universit{\"a}t T{\"u}bingen, Sand 1, 72076 T{\"u}bingen, Germany}

   \date{Received xxx; accepted xxx}

 
  \abstract
   {The high-energy environment of the host stars could be deleterious for their planets. It is crucial to ascertain this contextual information to fully characterize the atmospheres of terrestrial exoplanets.}
   {We aim to fully characterize a unique triple system, LTT1445, with three known rocky exoplanets around LTT 1445A.}
   {The X-ray irradiation and flaring of this system are studied through a new 50 ks Chandra observation, which is divided into 10 ks, 10 ks, and 30 ks segments conducted two days apart, and two months apart, respectively. This is complemented by an archival Chandra observation approximately one year earlier and repeated observations with eROSITA (extended ROentgen Survey with an Imaging Telescope Array), the soft X-ray instrument on the Spectrum-Roentgen-Gamma (SRG) mission, enabling the investigation of X-ray flux behavior across multiple time scales. With the observed X-ray flux from the exoplanet host star A, we estimate the photo-evaporation mass loss of each exoplanet. With the planet modeling package, \texttt{VPLanet}, we predict the evolution and anticipated current atmospheric conditions.}
   {Our Chandra observations indicate LTT 1445C as the dominant X-ray source, with additional contribution from LTT 1445B. LTT 1445A, a slowly-rotating star, exhibits no significant flare activity in the new Chandra dataset. Comparing the flux incident on the exoplanets, LTT 1445BC components do not pose a greater threat to the planets orbiting LTT 1445A than the emission from A itself. According to the results from the simulation, LTT 1445Ad might have the capacity to retain its water surface.}
   
   {}

   \titlerunning{Variability of LTT1445ABC}
   \authorrunning{S. Rukdee et al.}
   \maketitle

%

\section{Introduction}

Exoplanet science is entering a new phase focused on characterizing exoplanet atmospheres and identifying potentially habitable planets. This goal is currently attainable through spectroscopic analysis of nearby terrestrial exoplanets around mid to late M dwarfs, particularly those with slow rotation, relatively low activity, and masses ranging from 0.10 to 0.25 Earth masses \citep{Morley2017, NAP25187}. However, caution must be exercised when characterizing potentially habitable planets due to the potential impact of the host star's high-energy environment on their atmospheres.

Exoplanets orbiting M dwarfs are prime targets for characterizing the habitability of other worlds \citep{Shields2016}. This is because the host stars are abundant \citep{Bochanski2010}, and the habitable zone (HZ) is close to the star, making discovery and characterization possible on short time-scales \citep{Kasting1993, Kopparapu2013}. Population studies \citep[e.g.,][]{dressing_charbonneau2015} estimate that, on average, for every seven M dwarfs, the orbit of at least one Earth-size planet is in the HZ. So far, numerous exoplanets with a rocky composition have been detected. A significant proportion of these rocky planets is found near M-type stars \citep{Shields2016}. However, the proximity of these planets to their host stars can pose potential hazards to their physical and chemical characteristics. Exoplanets close to M dwarfs (and also F, G and K stars) can be exposed to high levels of X-ray and extreme ultraviolet radiation (XUV) and strong particle fluxes from stellar winds or coronal mass ejections \citep{Lammer2011}. On average, two-thirds of these cool stars are active, according to a survey of 238 nearby M dwarfs from Earth \citep{West2015}. Many studies consider XUV radiation from M dwarfs as deleterious for life \citep{Heath1999, Tarter2007, Lammer2009, Shields2016, Meadows2018}.

To assess habitability and comprehend the planetary atmospheric escape mechanisms contributing to planetary mass loss and oxygen accumulation, it is crucial to analyze the high-energy environment of M dwarfs. This involves examining stellar activity, variability, and flare frequency. For comparison, while the Sun emits the most energetic flares at an energy level of $\mathrm{10^{32}\,erg}$ once per solar cycle \citep{youngblood2017}, M dwarfs can experience flares of the same energy level every day \citep{Audard2000}. Moreover, even flares from inactive stars could affect the chemistry of the atmospheres of orbiting planets \citep{Hawley2014}.

The XUV activity is related to the age of the star and its rotation. In this context, \cite{Skumanich1972}, established a link between a star's rotation rate and its age, stating that the equatorial speed of a star decreases as the inverse square root of its age within the range of 100 Myr to 10 Gyr. \cite{Pallavicini1981} later identified a correlation between rotation and activity, expressing X-ray brightness as $L_x(v \sin i)^{1.9}$, providing initial evidence of dynamo-induced stellar coronal activity. For fast rotators, this relationship saturates at $\mathrm{L_x/L_{bol}=10^{-3}}$ \citep{Micela1985}. The X-ray-rotation relation for slow rotators is sparsely populated, posing challenges in accurately determining its turnover point and slope. \cite{Stelzer2016} used X-ray data to incorporate UV emission as a diagnostic for chromospheric activity in field M stars, deriving a saturation level of $\mathrm{LogL_{x,sat} = 28.6\pm0.3 \,erg/s}$ for M3-M4 stars with $\mathrm{P_{rot}}$ < 10 days. \cite{Wright_2011} compiled photometric rotation periods for 824 late-type stars, and computed the Rossby number \citep[Ro;][]{Noyes1984} as an evaluation of the convective turnover time. The Rossby number has been used to probe chromospheric activity in relation to rotation and convection, and applied in the context of X-ray observations \citep{Walter_Bowyer_1981, Noyes1984, Micela1985, Dobson_Radick_1989, Pizzolato2003}. Two regimes have been defined, with a saturated regime at Rossby number below 0.13, and an unsaturated regime presenting a steep decline of the X-ray-to-bolometric luminosity ratio at higher Rossby numbers. \cite{Wright2018} later expanded the sample to include 19 slowly rotating fully convective stars, suggesting that both fully- and partly- convective stars operate similar dynamos, relying on the interplay of rotation and turbulent convection.

\begin{table}[h]
\begin{threeparttable}
    \centering
    \caption{The three stars in the LTT 1445 system}
    \begin{tabular}{ccccc}
    \hline\hline
         Star & Spectral  & $M_*^{b}$ &  $R_*^{b}$  & $L_{bol}^{b}$ \\
         & Type$^{a,b}$ & [$M_{\astrosun}$] & [$R_{\astrosun}$] &  [$10^{31}\, erg/s$]     \\
         \hline
        A & M 3 & 0.257$\pm$0.014  & 0.268$\pm$0.027 & 3.04 \\

        B & M 3.5  & 0.215 $\pm$ 0.014 & 0.236$\pm$0.027 & 2.28  \\ 

         C & M 4 & 0.161 $\pm$ 0.014 & 0.197$\pm$0.027 & 1.41 \\ 
    \hline     
    \end{tabular}
    \label{tab:stars}
    \begin{tablenotes}
      \small
      \item  {Reference $^a$ \cite{Reid2004}; $^b$ \cite{Winters2019}  }.
    \end{tablenotes}
\end{threeparttable}
\end{table}

The closest transiting exoplanetary system

to us is LTT 1445, a triplet star system with three rocky exoplanets \citep{Winters2019, Winters2022, Lavie2022} around the A component. The LTT 1445 system, located 6.9 parsecs away from the Sun, consists of three low-mass M-type stars: A, and a binary pair, B and C. This stellar system denoted as ABC, has parameters shown in Table \ref{tab:stars}. According to \cite{Reid2004} and \cite{Winters2019}, the spectral type of star A in the LTT 1445 system is classified as M3. Stars B and C are classified as M3.5 and M4, respectively. The separation between the components LTT 1445A and LTT 1445BC in this astrometric system is currently about 34 astronomical units (AU) or approximately 7" on the sky \citep{Winters2019, Brown2022}, with an orbital period of 250 years. The B and C components orbit each other with a period of 36 years, and are separated by 1.25" on the sky \citep{Brown2022}. Ground-based MEarth observations have shown in the image that the BC pair is always blended, while the blending between A and BC is variable \citep{Winters2022}. 

Previous observations have documented stellar activity in the LTT 1445 system. The TESS light curve revealed stellar flares and rotational modulation attributed to star spots, likely originating from either or both of the B or C components \citep{Winters2019}. A super-flare with an energy of $\mathrm{10^{34}\, erg}$ was also observed in TESS data where the triple system is not resolved \citep{Howard2019}. Subsequently, the system was detected by eROSITA in the first eROSITA all-sky survey (eRASS1), where it was assumed that all the detected flux originated from the known active stars LTT\,1445\,BC \citep{Foster2022}. As expected for active stars, \cite{Brown2022} revealed a wide range of variability in the A component from the 30ks Chandra observation conducted in 2021. This suggests highly fluctuating and stochastic high-energy irradiance on its planets, leading to potential atmospheric alterations. Leveraging the high spatial resolution of Chandra, \cite{Brown2022} resolved all three components and reported the X-ray flux from the A flare at a level of $3.61\pm0.27 \times \mathrm{10^{-13}\, erg\,cm^{-2}s^{-1}}$, while the BC flare emitted $20.2\pm0.9 \times \mathrm{10^{-13}\, erg \,cm^{-2}s^{-1}}$. The quiescent X-ray luminosity for the A component was reported at the low level of $3.7 \times \mathrm{10^{25}\, erg\, s^{-1}}$.

The proximity and brightness of its stars makes the LTT1445 system stand out as a promising laboratory for in-depth planet characterization. In the literature, XUV flux around LTT 1445, or PMI03018-1635, was explored by Stelzer2013, who combined archival ROSAT, XMM–Newton, and GALEX data for the BC component, measuring the X-ray flux in the 0.2-2.0 keV energy band with $\mathrm{\log F_x \, (mW/m^2)}$ equal to -12.06 and $\mathrm{\log F_{NUV} \, (mW/m^2)}$ \sredit{and $\mathrm{\log F_{FUV} \, (mW/m^2)}$ are -12.46 and -13.07 respectively.} Despite these insights, several questions remain from \sredit{previous studies}\cite{Brown2022}, including whether LTT 1445B \sredit{happened to be previously} observed during an unusually X-ray quiescent time and the potential influence of BC activity on planets around the A component.

\begin{figure}[h!]
    \centering
    \includegraphics[width=\columnwidth]{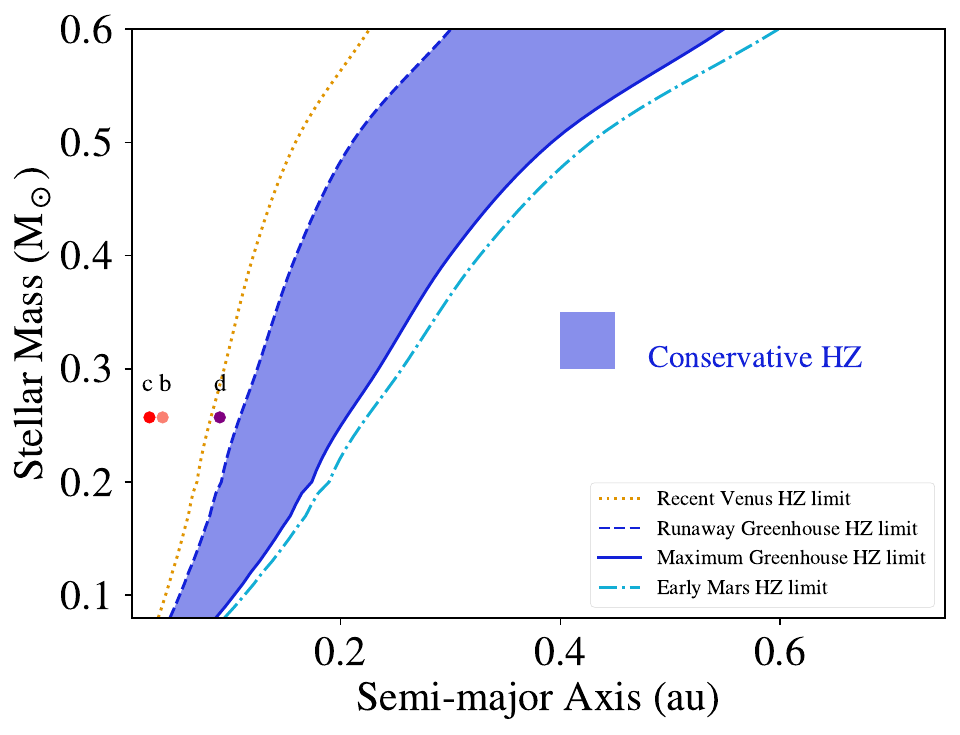}
    \sredit{\caption{Positions of the LTT-1445A planets (colored circles) with respect to the habitable zone \citep[blue;][] {Kopparapu2013} where water is liquid. Planet LTT1445Ad is near the conservative Habitable Zone. Yellow dotted line indicates a limit of the Recent Venus HZ. Between the cyan dash-dotted line and the conservative HZ zone, water would be frozen. White areas outside all the HZ limits are not expected to have liquid water.}}

    \label{fig:habitable_zone}
\end{figure}


\section*{Rocky Planets in \sredit{the} Triple Star System}

The star LTT 1445A hosts three rocky exoplanets. The detection of the initial two planets involved the analysis of photometric data from various sources, including MEarth, Las Cumbres Observatory Global Telescope (LCOGT), and Transiting Exoplanet Survey Satellite (TESS), to detect transits. To further characterize these exoplanets, spectroscopic data from 136 radial velocities (RV) of LTT 1445A were collected over two years using high-resolution spectrographs. 

Detailed information about the planetary parameters of the LTT 1445 system can be found in Table~\ref{tab:companions}. \cite{Winters2022} precisely measured the masses of both planets, and the mass-radius relation, based on the interior structure model \citep{Zeng_Sasselov2013}, indicated an Earth-like composition with approximately 33\% iron and 67\% rock. Subsequently, \cite{Lavie2022} utilized 84 the Echelle SPectrograph for Rocky Exoplanets and Stable Spectroscopic Observations (ESPRESSO) high-resolution spectra and nested sampling algorithm \cite{Buchner2014}, a reliable statistical method \cite{Nelson2020} \sredit{for} discovering LTT 1445Ad, which orbits at the inner boundary of the habitable zone of its host. Here, we estimate the radius of this planet to be 1.29 $R_\earth$ based on the mass-radius prediction in Fig.~3 of \cite{Chen_Kipping2017}. \sredit{Fig. \ref{fig:habitable_zone} places the orbital distances of the three planets in the context of the habitable zone with suggested limits from \cite{Kopparapu2013}. Here we assumed a stellar age of 2 Gyr, which we discussed this assumption in Section \ref{sec:discussion}}. 

The Habitable Zone is the distance from the star where a liquid water layer can be sustained. The model proposed by \cite{Kopparapu2013} assesses the habitability potential of extrasolar terrestrial planets by using of stellar flux incident on a planet rather than equilibrium temperature. This eliminates the dependency on planetary albedo, which varies based on the spectral type of the host star.

\begin{table}[ht!]
    \caption{The three rocky exoplanet companions of LTT~1445A}
    \begin{tabular}{cccccc}
    \hline\hline
         Planet & $M_P$ &  $R_P$ &  a & $P $ \\
        & [$M_{\earth}$] & [$R_{\earth}$] &  [AU] & [days] \\ 
         \hline

        c$^d$ & $1.54^\dscript{+0.20}{-0.19}$ & $> 1.147^\dscript{+0.055}{-0.0054}$ & $0.02656$ & 3.12390 \\ 
        
        b$^c$ & $2.87^\dscript{+0.26}{-0.25}$  & $1.305^\dscript{+0.067}{-0.060}$ & $0.03807$ & 5.35877 \\ 
           
         d$^e$ & $2.72 \pm 0.75$ & non-transiting* & 0.09 & 24.30388 \\ 
    \hline
    \end{tabular}
    \begin{tablenotes}
    \small
      \item  {Reference $^c$ \cite{Winters2022}; $^d$ \cite{Winters2019}; $^e$ \cite{Lavie2022}  }, *estimated \sredit{as} 1.29$R_\earth$ \sredit{(see text)}.
    \end{tablenotes}
    \label{tab:companions}
\end{table}

This study revisits the LTT 1445 triplet star system with new Chandra and eROSITA data\sredit{. The X-ray observations collected over 3 years are described in Section \ref{sec:observations}. 
The analysis methods are described in Section \ref{sec:method}. In Section \ref{sec:results}, we examine the variability in X-ray flux emitted by the individual stars. Additionally, we estimate various parameters related to the activity level, mass loss rate, and mass loss evolution of each known planet within the system. In Section \ref{sec:discussion}, we study oxygen build-up caused by XUV irraditation} and discuss its implications within the context of LTT 1445. The summary is provided in Section \ref{sec:summary}.

\section{Observations} \label{sec:observations}
\subsection{Chandra Observations} \label{subsec:Chandra_obs}

\begin{figure*}
    \centering
    \includegraphics[width=\textwidth]{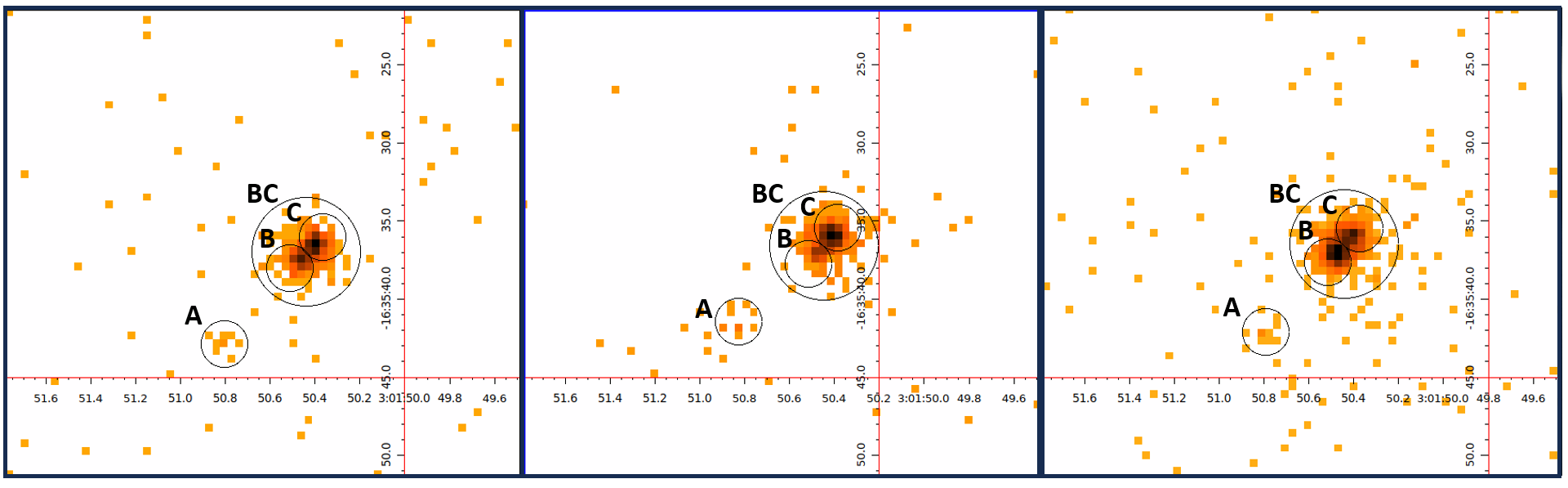}
        \caption{Chandra/ACIS image of the LTT~1445 system for the three observations. North is up, east is left. The black circles show the photon extraction region from the three Chandra observations: 2022-10-10, 2022-10-12 and 2022-11-24 (left to right) for each component A, B, C, and BC together. 
        }
    \label{fig:chandra_images}
\end{figure*}

Chandra ACIS-S observations, totaling 50 ks, were granted as part of MPE GTO time (proposal 23100661 cycle 23). This system revisit occurred approximately 1.5 years after the Chandra observation reported by \cite{Brown2022}. Our observations were divided into three epochs. The first and second observations were taken 2 days apart, while the last observation occurred 1.5 months after the first one. The observation log is detailed in Table~\ref{tab:Chandra_obs_log}. Benefiting from Chandra's high spatial resolution, we are able to analyse the X-ray emission from the components of LTT1445, namely A and BC, separately as shown in Fig.\ref{fig:chandra_images}. The three stellar components are aligned in a northeast-to-southwest direction, with the projected distance from A to B being approximately 7".

\begin{table}[h]
    \centering
    \caption{Chandra Observation log}
    \begin{tabular}{cccc}
    \hline\hline
         Obs ID &  Instrument  & Dur.(s) &  Obs.time \\
         \hline
        25993 & Chandra ACIS & 11000  & 2022-10-10T09:08:22  \\
        27476 & Chandra ACIS & 10000  & 2022-10-12T08:38:48  \\
        27477 & Chandra ACIS & 29000  & 2022-11-22T23:38:24  \\
    \hline
    \end{tabular}
    \label{tab:Chandra_obs_log}
\end{table}

\begin{figure*}
    \centering
    \includegraphics[height=6.8cm]{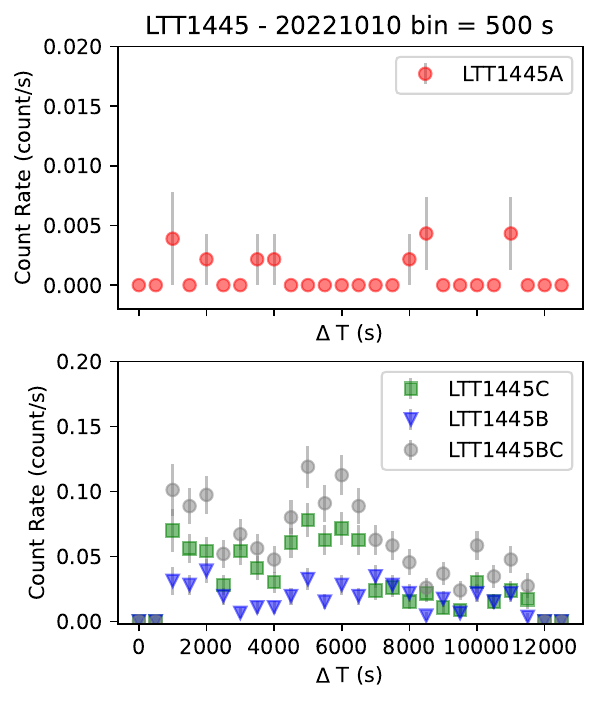}
    \includegraphics[height=6.8cm]{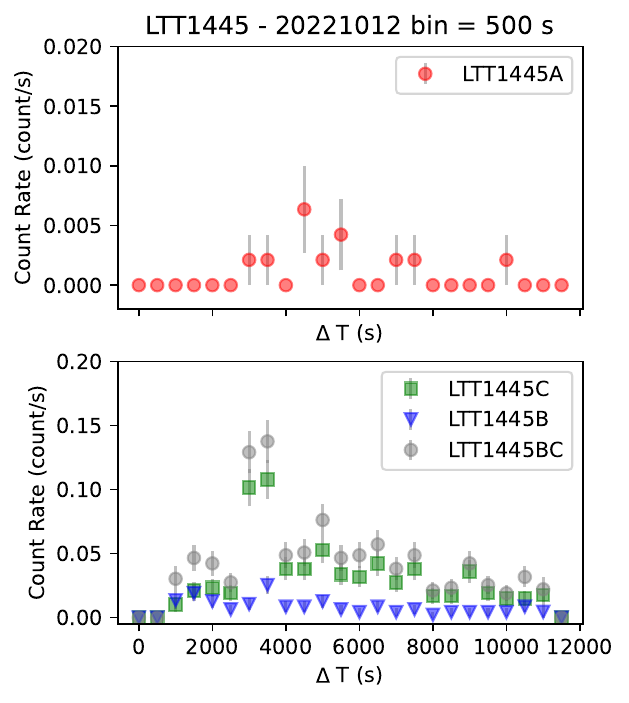}
    \includegraphics[height=6.8cm]{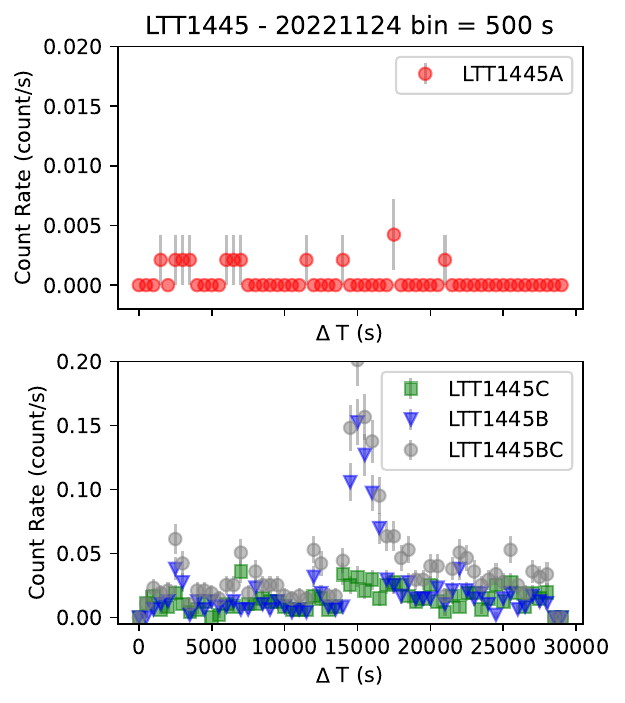}

    \caption{Light curves extracted from the three Chandra observations: 2022-10-10, 2022-10-12 and 2022-11-24 (left to right) of the A component (red, top panels), B in green square, C in blue triangle and combined BC in gray diamond (bottom panels).}
    \label{fig:chandra_lightcurve}
\end{figure*}

The data were analyzed following standard procedures outlined in the CIAO package version 4.14, as detailed by \cite{Fruscione2006}. In our analysis, the event list was filtered to the 0.3-10 keV energy range. Circular regions with a radius of 1.5" were applied \sredit{(see Fig.~\ref{fig:chandra_images})} for source data extraction for each of the three components: LTT 1445 A, B, and C. Larger circular regions with a radius of 3" were employed for the combined BC components, and the background region was defined as an annulus. We jointly extract data from BC to analyze the flux and assess potential impacts on the planets orbiting the A component. Additionally, we extract \sredit{data for} each star individually, to discern the activity level of each star. 

\sredit{In the region corresponding to the A component in Fig. \ref{fig:chandra_images}, \sredit{9, 11, and 12 counts were detected in the three observation listed in Table \ref{tab:Chandra_obs_log}}. For the region of B, we recorded 194, 81, and 598 \sredit{counts} for each respective observation. Regarding C's region, \sredit{381, 335, and 391 counts were detected} for each respective observation.}

After identifying the source locations, we extracted the light curve from each component. The light curves, with a bin size of 500 seconds for each dataset, are presented in Fig.~\ref{fig:chandra_lightcurve}. \sredit{LTT\,1445\,A showed relatively little variability} across all three datasets, as indicated by the top panel in red at a level of 0.005 counts/second. For LTT1445BC, the initial two datasets on the left were captured within two days. Both components exhibit a comparable activity level; B is in a quiet period with an average of 0.03 counts/second, while C is considered quasi-quiescent (qq) or elevated at a level of 0.05 counts/second. The bottom-left panel does not display prominent flare characteristics; instead, it exhibits quasi-quiescent behavior. \sredit{Table \ref{tab:flare_intervals} lists the average count rates (recorded number counts divided by effective exposure time) for the various light curve segments of BC.}

In the middle panel of Fig.~\ref{fig:chandra_lightcurve}, a flare event was observed in component C at a level of 0.10 counts/second. The last dataset, collected approximately 1.5 months later, distinctly demonstrates the variability of this system, with a flare predominantly originating from component B at a level of 0.15 counts/second. 

We study the stellar spectrum of each component during both quiescence and flaring. \sredit{As the A component remained quiet in all datasets, we consider the entire data set as one, quasi-quiescent state. For BC, as described above and listed in Table~\ref{tab:flare_intervals}, the X-ray light curves were classified into quiescent, qq or flare periods. For the A component, we classify all our new Chandra data as quiescent. Based on these time intervals, we extracted the spectrum of each component to compare spectral changes over activity levels.}

For the new 50 ks Chandra observation, our spectrum extraction followed the standard \sredit{procedures} from \texttt{CIAO}. \sredit{In addition to the 
regions of interest for the three stars (Fig. \ref{fig:chandra_images}), an annular background region was defined,} sufficiently large to encompass the background around the source ABC. The \texttt{specextract} script was applied to extract spectra using the time intervals identified above, with the regions already used above. \sredit{These spectra are analysed below in section~\ref{sec:method}}

\begin{table}[h!]
    \centering
    \caption{Flare interval identification for BC components from the Chandra observation in 2022}
    \begin{tabular}{cccc}
    \hline\hline
          Obs.date  &  Intervals & Duration & Count rate \\
          &  &  [s] & [cnt/s] \\
         \hline
        2022-10-10 & qq & 0-12000 & 0.055
    \\
        2022-10-12 & Quiet & 0-1800 & 0.019 \\
        & Flare & 1800-4000 & 0.084\\
        & qq & 4000-12000 & 0.038
   \\
        2022-11-24 & Quiet & 0-14000 & 0.025 \\ 
        & Flare & 14000-18000 & 0.114 \\
        & qq & 18000-30000 & 0.032 \\
 \hline
    \end{tabular}
    \label{tab:flare_intervals}
\end{table}

\subsection{eROSITA Observations}
The eROSITA all-sky survey \citep[eRASS;][]{Predehl2021}, started in 2019, and passed over LTT1445 four times: eRASS1 \citep{Merloni_2024} on 2020-01-25, eRASS2 on 2020-07-26, eRASS3 on 2021-01-19, eRASS4 on 2021-07-27), with gaps of 6 months. The initial Chandra observation in June 2021 \citep{Brown2022} lies between eRASS3 and eRASS4. 

In this study, we leverage sources detected in the 0.2–2.3 keV range from eRASS1 to eRASS4 for continuous system monitoring. Specifically, we use the eRASS1 to eRASS4 source catalog versions 221031, 230619, and 230119 processed by the \texttt{eSASS} pipeline version 020 \citep{Brunner2022}. 

\section{Methods} \label{sec:method}
\sredit{In this section, we describe the adopted statistical approach \sredit{for spectral fitting} (Sect \ref{subsec:stat_method}), the spectral model in Sect \ref{subsec:spectral_model}, and the model for X-ray driven evaporation \sredit{of the planet atmospheres} in Sect \ref{subsec:atmo_model}.}

\subsection{Statistical Method} \label{subsec:stat_method}

In X-ray astronomy, the chi-square statistic ($\mathrm{\chi^2}$) and C-statistic (Cstat) \citep{Cash1979} are commonly used in spectral fitting. These differ in assumptions and applications. The $\mathrm{\chi^2}$ is rooted in Gaussian statistics, suitable for normally distributed data uncertainties, quantifying the fit's goodness by comparing observed and expected values. In contrast, the C-stat is ideal for Poisson statistics, common in X-ray data. Particularly effective in low count rates or background-dominated scenarios, the C-stat provides a more accurate representation of data statistics \citep{Wheaton1995, Nousek1989}. Unlike the $\mathrm{\chi^2}$, studies show that the C-statistic in X-ray spectra analysis yields unbiased estimates of model parameters and their uncertainties \citep{Kaastra2017, Buchner2022}. In contrast, the $\mathrm{\chi^2}$ statistic, even with uncertainties, can lead to biased model parameter estimates, especially with fewer than 40 counts \citep{Humphrey2009}—see also Figure 6 in \cite{Buchner2022}. Traditionally, the analysis of the X-ray plasma emission model has employed the chi-square statistic. However, in this work, we conduct our analysis using the C-stat and Bayesian methods, comparing the results with traditional chi-square analysis.

X-ray spectra are analysed with the spectral fitting package Bayesian X-ray Analysis, \texttt{BXA} \citep{Buchner2014}. \texttt{BXA} integrates the nested sampling algorithm \texttt{UltraNest} \citep{Buchner2021} with the fitting environment \texttt{CIAO/Sherpa} \citep{Fruscione2006}. This software is particularly advantageous as it provides precise parameter constraints, even in scenarios with very low counts or intricate parameter spaces characterized by significant degeneracy. Moreover, \texttt{BXA} requires minimal user input for initializing parameter values before initiating systematic fitting.

\begin{figure}[h!]
    \centering
    \includegraphics[width=0.8\columnwidth]{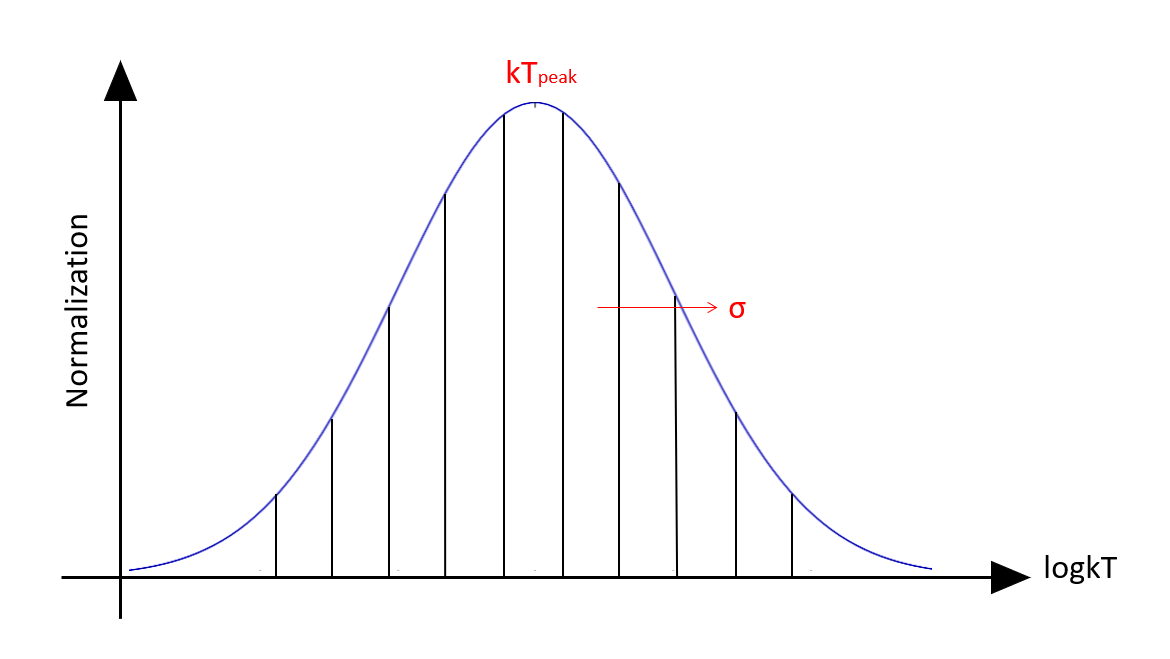}
        \caption{An illustration of the log-Gaussian 
        temperature distribution model based on a fixed grid of single-temperature components. These parameters include \sredit{the normalization of the gaussian peak of the temperature distribution (normpeak),} the temperature of the peak (kTpeak), and the standard deviation of the distribution (sigma). Based on these parameters, the normalization is assigned following the Gaussian formula at each temperature (T) for the 10 APEC components.}
    \label{fig:temperature_distribution}
\end{figure}

\subsection{Spectral Model} \label{subsec:spectral_model}

For the modeling process, we employed the \texttt{APEC} model \citep{Smith2001}. As a compromise between realistic modeling and handling moderate photon counts, we adopted a plasma with a log-Gaussian temperature distribution. Typically, a single-temperature plasma (with 2 free parameters: normalization and temperature) is assumed for relatively low-count spectra. The plasma is modeled using the Astrophysical Plasma Emission Code (\texttt{APEC model}), which has all metal abundances linked. High-quality M dwarf spectra show that the plasma temperature distribution is smooth and broad \citep{Robrade2005}. The temperature distributions resemble a log-Gaussian during quiescent phases, and also during flares, which typically exhibit a broader and hotter temperature distribution \citep{Robrade2005}. Therefore, we model the spectra with a plasma featuring a log-Gaussian temperature distribution. \sredit{This approach is expected to capture the behavior of the plasma temperature better than a single temperature or two temperature plasma model (kT1 and kT2).} In practice, this is achieved by an ensemble of 10 APEC components on a logarithmic temperature grid, where the normalizations follow a bell curve as part of the modeling process in \texttt{BXA} \footnote{\url{https://github.com/SurangkhanaRukdee/BXA-Plasma}}. \sredit{The Gaussian temperature distribution is therefore approximated by summing many single-temperature components. We assume ten components as a trade-off between computational cost and quality of the approximation.} This is illustrated in Fig.~\ref{fig:temperature_distribution}. The free parameters are the peak temperature, peak normalization, and the width $\sigma$ of the log-Gaussian distribution. The list of priors is detailed in Table~\ref{tab:prior}. These parameters control the normalization of each component, as depicted in Fig. \ref{fig:temperature_distribution}. \sredit{In addition, the abundance parameter sets the metal abundances of all \texttt{APEC} model components.}.

\begin{table}[h]
    \centering
    \caption{Priors set for the APEC model}
    \begin{tabular}{llc}
    \hline\hline
        Parameter & Prior & Range\\
        \hline
        \sredit{abundance} & uniform & 0.0 - 1.0\\
        kTpeak & log-uniform & 0.1 - 5.0\\
        normpeak & log-uniform & $10^{-6}$ - 0.01\\
        sigma & uniform & 0.0 - 2.0\\
    \hline
    \end{tabular}
    \label{tab:prior}
\end{table}

In addition, \sredit{we model the background spectrum. By providing the analysis with additional information about the instrument and expected smooth correlations between data bins, more information can be extracted from low-count data. We adopt the Chandra background model of \cite{Simmonds2018} obtained through principal components analysis (PCA), and fit this model to each background spectrum with BXA. Throughout, our} analysis employs Poisson (C-Stat) statistics to jointly assess the background and source spectra.

\subsection{Modeling of Atmospheric Escape} \label{subsec:atmo_model}

To assess the impact of X-ray radiation on the planet, we use the \texttt{VPLanet} model \citep{Barnes2020}. \texttt{VPLanet} conducts extensive simulations of planetary system evolution, spanning Gyr timescales, with a primary emphasis on studying habitable planets. Its modules encompass various aspects such as internal, atmospheric, rotational, orbital, stellar, and galactic processes. These modules can be interconnected to enable the concurrent simulation of the evolution of terrestrial planets, gaseous planets, and stars. The code's validity is established through its capacity to replicate a range of observations and previously obtained results. In this work, we use some modules to estimate the influence of the observed XUV flux and predict water loss in the system. The simulation of atmospheric escape is conducted using the \texttt{AtmEsc}, \texttt{STELLAR}, and \texttt{FLARE} modules within the \texttt{VPLanet} framework. The \texttt{AtmEsc} module is designed to simulate the escape of planetary atmospheres and the release of surface volatiles based on energy-limited and diffusion-limited mechanisms. The VPLanet model has been developed based on previous research, indicating that for small planets, the escape rate correlates with the stellar XUV flux and inversely with the gravitational potential energy of the gas \citep{Lopez2012,Lammer_2013, Owen_Wu_2013, Owen_Wu_2017}. Because of the limited knowledge of exoplanetary atmospheric structures, \texttt{VPLanet} employs the energy-limited approximation. All uncertainties concerning the escape process physics are encapsulated in the XUV escape efficiency ($\mathrm{\epsilon_{xuv}}$) \sredit{following \cite{Bolmont2017_waterloss} 1D radiation-hydrodynamic simulations of atmospheric loss. Figure 29 in \cite{Barnes2020} illustrates \sredit{the relationship of the XUV atmospheric escape efficiency of water with XUV flux received by the planet}, with 0.1 as \sredit{a representative} value. Several studies, including those by \cite{Lopez2012, Lammer_2013,  Owen_Wu_2013, Owen_Wu_2017}, suggest that for small planets, the escape rate is linked to the stellar XUV flux and inversely proportional to the gravitational potential energy of the gas. Therefore, the model assumes $\mathrm{\epsilon_{xuv}\sim}$ 0.1 as a reasonable median value.} We caution that the energy-limited formalism used in \texttt{AtmEsc} is an approximate description of atmospheric escape. It does not account for wavelength dependence in upper atmosphere heating, varying with composition and temperature structure, which is omitted from the model. The \texttt{STELLAR} module models key characteristics in low-mass stars ($\mathrm{M_{\star} \leq 1.4 M_{\astrosun}}$), including radius, stellar radius of gyration ($\mathrm{r_g}$), effective temperature, bolometric luminosity, XUV luminosity, and rotation rate. This simulation employs bicubic interpolation over mass and time, with the \cite{Baraffe2015} models based on solar metallicity stars. 

The \texttt{FLARE} modules \citep{Amaral2022} integrate the flare frequency distribution model proposed by \cite{Davenport2019} into the conventional \texttt{STELLAR} package. \sredit{Combining the results from \texttt{STELLAR} and \texttt{FLARE} modules reveals that in M dwarfs}, flares contribute approximately 10\% more XUV emission \sredit{than the baseline} quiescent stellar levels. Such high XUV levels in the flaring star can result in the removal of up to an additional two terrestrial oceans (TO) of surface water on Earth-like planets, compared to the non-flaring star.

In our simulations, we place the three planets with observed mass, radius, and orbital radius \citep{Winters2019, Winters2022, Lavie2022} around the star. The \sredit{evolutionary model is the one by \cite{Baraffe2015}.}

The initial configuration comprises 1.0 Terrestrial Ocean (TO), equivalent to the amount of water on Earth. Once the Hydrogen envelope is eliminated, XUV photons trigger the dissociation of water molecules, resulting in increased Hydrogen escape and oxygen buildup in the atmosphere. The water escape is modeled based on \cite{Bolmont2017_waterloss}, taking into account the habitable zone limit from \cite{Kopparapu2013}.

The examination of stellar evolution follows the trajectory of a stellar model described in \cite{Baraffe2015}. This considers factors such as magnetic braking \citep{Reiners2014} and the XUV evolution model. The simulation input parameters are based on the study of solar-type stars during their quiescent periods \citep{Ribas2005}. 

The \texttt{FLARE} model in Fig.~\ref{fig:StellarEvol} is detailed in \cite{Amaral2022}, incorporating a relation between flare energy and frequency \citep{Davenport2019}, dependent on the \sredit{stellar} type and age. The \texttt{FLARE} module computes the average XUV for each simulation time-step, integrating from \sredit{a minimum to a maximum considered} flare energy. Generally, the inclusion of flares increases \sredit{the} XUV luminosity by 10\%. In our work, we assumed flare energies between $\mathrm{10^{30}}$ and $\mathrm{10^{33}}$ erg following \cite{Davenport2019}. \sredit{That} study found that flare activity decreases as low-mass stars spin down and age. Additionally, the \sredit{flare} frequency 
distributions from observed stars in the Kepler field show no significant change in the power law slope with age. It is noteworthy that the \cite{Davenport2019} model over-predicts the superflare rate as it was constructed from younger and more active stars than the Sun, of which approximately 3\% of the catalog samples \citep{Davenport2016} are M-stars.

Finally, \sredit{we explore planetary mass loss due to the host star. This atmospheric escape, also known as atmospheric evaporation,} occurs when high-energy radiation and charged particles from the host star ionize the gas molecules in the planet's upper atmosphere, causing them to escape into space. The rate of atmospheric escape depends on several factors, including the planet's mass, radius, composition, magnetic field, and distance from the host star. Notably, \cite{Ketzer_Poppenhaeger2023} found that the natural scatter of stars in their spin-down evolution can have a significant impact on the accumulated mass loss of exoplanets. We determined the mass loss rate for the two transiting planets where the current planetary radius $R_{pl}$ was observed, Ab and Ac—using the following equation:

\begin{equation}
\dot{M} = \epsilon\times \frac{\pi R_{pl}F_{XUV}}{GM_{pl}/R_{pl}}
\end{equation}

\noindent
Here, $\dot{M}$ represents the mass loss rate, $\epsilon$ denotes the efficiency from the mass loss model \citep{Lopez2012, Owen2012, Foster2022}, $R_{pl}$ stands for the planet radius, $M_{pl}$ for the planet mass, $G$ is the gravitational constant, and $F_{\rm XUV}$ combines the observed X-ray flux and the estimated EUV flux from the host star. The EUV luminosity of the star was derived from X-ray luminosity using Eq.~3 in \cite{SanzForcada2011}. 

\sredit{Next, we calculate Roche lobe overflow, thermal escape, and radiation-recombination-limited processes of an atmosphere with the \texttt{AtmEsc} module \citep{Barnes2020}.} This incorporates water photolysis, Hydrogen escape, oxygen escape, and oxygen build-up. The thermal escape is approximated with an energy-limited formula. The configuration parameters for \texttt{VPLanet} are reported in Table \ref{tab:vplanet_param}. Detailed information on the stars' and planets' parameters is provided in Tables \ref{tab:stars} and \ref{tab:companions}, respectively \citep{Winters2019,Winters2022,Lavie2022}. The planetary characteristics in our simulation are modeled after Proxima Centauri b \citep{Barnes2016}. The \sredit{initial envelope mass} of the planet is set \sredit{to} 0.001 $M_\earth$. Our assumption includes an initial water abundance of 1.0 Terrestrial Oceans (TO) for the system. We start our simulation at a stellar age of \sredit{5 Myr where we assume the planetary disk of these low-mass stars disperses \citep{Pfalzner2022}. This conservative assumption presumes that the planet forms early in the disk evolution and remains irradiated by the star for the maximum possible duration.} For the flare energy, we follow \cite{Davenport2019}, which suggests Kepler (optical) flare energies of $\mathrm{10^{33}}$ and $\mathrm{10^{36}}$\, erg. 


\begin{table}[h!]
    \centering
    \caption{Parameter of Star and Planets used for the simulation with \texttt{VPLanet} for LTT1445A}
    \begin{tabular}{lc}
    \hline\hline
         Parameter &  Value  \\
         \hline
         
        Envelope mass ($M_{\earth}$) & 0.001   \\
        Surface water (TO)  & 1.0   \\
        XUV water escape efficiency  & \cite{Bolmont2017_waterloss}    \\
        Thermosphere temperature (K)  & 400   \\
        Stellar mass ($M_{\astrosun}$) & 0.257   \\
        Saturated XUV luminosity fraction & $\mathrm{10^{-3}}$ \\
        Initial Age (Myr) & \sredit{5.0  }   \\
        Flare Energy (erg) & $\mathrm{10^{33}}$ to $\mathrm{10^{36}}$   \\
    \hline
    \end{tabular}
    \label{tab:vplanet_param}
\end{table}

\section{Results} \label{sec:results}

\begin{table*}[h!]
    \centering
    \caption{X-ray properties for LTT 1445A across observations}
    \begin{tabular}{cccccccc}
    \hline\hline
         Date & Observatory & Activity & Duration & $f_X $ & $\mathrm{logL_X} $ & $kT_{peak}$  & Ref.\\
          &  &  & [ks] & [$\mathrm{10^{-13}\, erg\,cm^{-2}\,s^{-1}}$] &  & [keV] &   \\
         \hline
        
         2021-06-05 & Chandra &  Flare & 6.66  & $3.61\pm0.27$ & $27.31\pm0.10$ & $1.02\pm0.10$  & Brown 2022 \\
         2021-06-05 & Chandra &  Elevated & 11.6 & $0.66\pm0.10$ & $26.57\pm0.07$ & $0.59\pm0.29$  & Brown 2022 \\
         2021-06-05 & Chandra &  Quiescent & 12.2 & $0.066\pm0.033$ & $25.57^\dscript{+0.18}{-0.30}$ & $1.02\pm0.10$ & Brown 2022 \\
         2022-10-10 & Chandra &  Quiescent & 12.0 & $0.259^\dscript{+0.12}{-0.08}$  & $26.16\pm0.24$  & $0.80\pm0.94$  & This work \\
         2022-10-12 & Chandra &  Quiescent & 12.0 & $0.19^\dscript{+0.15}{-0.07}$ & $26.0\pm0.4$& $0.68\pm0.87$ & This work \\
         2022-11-24 & Chandra &  Quiescent & 30.0  & $0.18^\dscript{+0.05}{-0.05}$ & $26.03\pm0.13$ & $0.94\pm0.76$  & This work \\
    \hline
    \end{tabular}
    
    \label{tab:LTT1445A_properties}
\end{table*}

\begin{figure*}[h!]
    \centering
    \includegraphics[height=5.5cm]{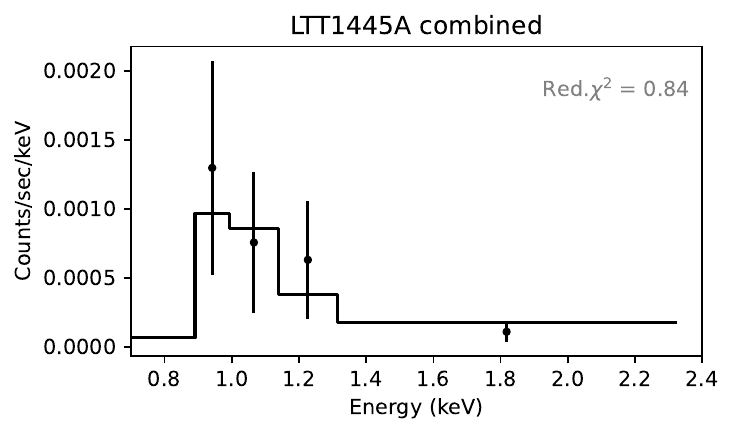}
        \includegraphics[height=5.5cm]{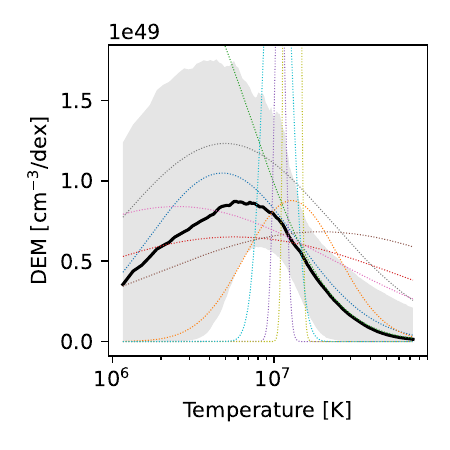}

    \includegraphics[height=5.5cm]{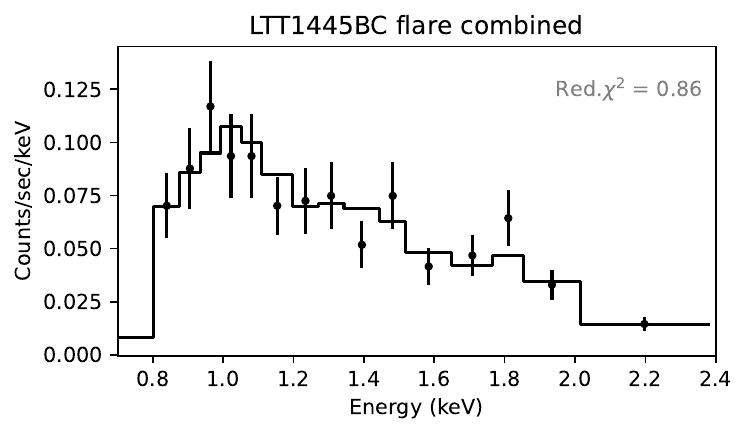}
            \includegraphics[height=5.5cm]{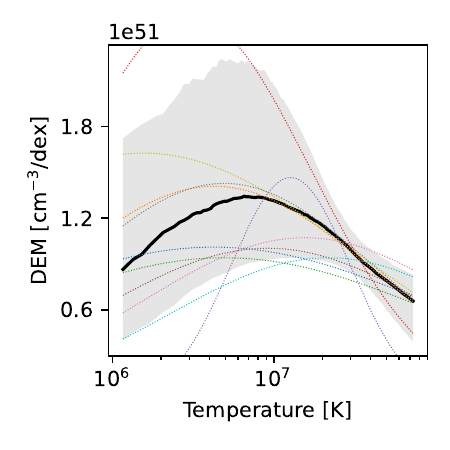}

    \includegraphics[height=5.5cm]{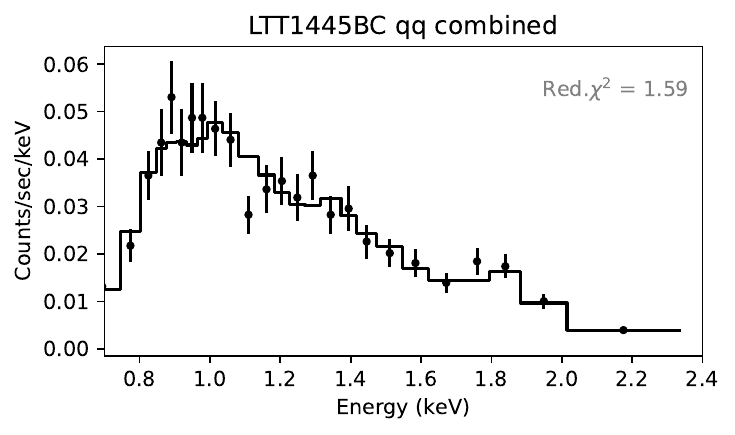}
            \includegraphics[height=5.5cm]{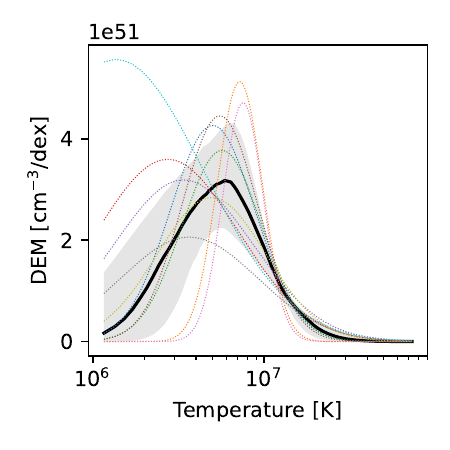}

    \caption{
    The left panels show the Chandra ACIS spectra of LTT 1445A combined (top panel) and LTT 1445BC combined Flare (middle) and Quasi-Quiescent (bottom). The corresponding right panels show the reconstructed continuous Differential Emission Measure (DEM) distribution (approximated with a grid during the fit, see Fig.\ref{fig:temperature_distribution}). The black curve shows the median of the posterior DEM predictions, and the gray band contains 68\% of the distribution. The colorful dotted lines are ten random posterior samples of possible DEM.  The top right panel y-axis has a 100 times lower normalization than the lower right panels. The bottom right panel DEM has a narrow shape plummeting at  $2\times10^7\,\mathrm{K}$, while the middle right panel shows a broad distribution. } 
    
        \label{fig:LTT1445ABC_spec}
\end{figure*}

\begin{table*}[h!]
    \begin{center}
    \caption{X-ray properties for LTT 1445BC across observations}
    \begin{tabular}{cccccccc}
    \hline\hline
         Date & Observatory & Activity & Duration & $f_X $ & $\mathrm{logL_X} $ & $kT_{peak}$  & Ref.\\
          &  &  & [ks] & [$\mathrm{10^{-13}\, erg\,cm^{-2}\,s^{-1}}$] &  & [keV] &   \\
         \hline
         2021-06-05 & Chandra & Flare BC & 5.2  & $20.2\pm0.90$ & $28.06\pm0.02$ & $0.38\pm0.1$*  & Brown 2022 \\
          2021-06-05 & Chandra & Nonflare BC & 23.4 & $7.3\pm0.20$ & $27.61\pm0.01$ & $0.76\pm0.03$  & Brown 2022 \\

          2022 all & Chandra & Flare BC & 6.20 & $14.9^\dscript{+0.93}{-0.95}$**  & $27.933\pm0.027$  & $0.81\pm0.8$  & This work \\
          2022 all & Chandra & Nonflare BC & 47.8 & $7.63^\dscript{+0.34}{-0.34}$** & $27.64\pm0.019$& $0.43\pm0.13$ & This work \\
          2022-10-12 & Chandra & C - Full & 12.0 & $10.86^\dscript{+1.48}{-1.09}$  & $27.80\pm0.05$  & $0.33\pm0.25$  & This work \\
          2022-11-24 & Chandra & B - Full & 30.0 & $5.88^\dscript{+0.48}{-0.59}$  & $27.53\pm0.04$  & $0.82\pm0.67$  & This work \\
         
    \hline
    \end{tabular}
    \end{center}
    \tiny{*The lower temperature ($kT_1$) from the two temperature fit\\
    **Flux calculated for energy band 0.6-2.3 keV to match the eROSITA flux}

    \label{tab:ltt1445bc_properties}
\end{table*}

\begin{figure*}[h!]
    \centering
    \includegraphics[width=\textwidth]{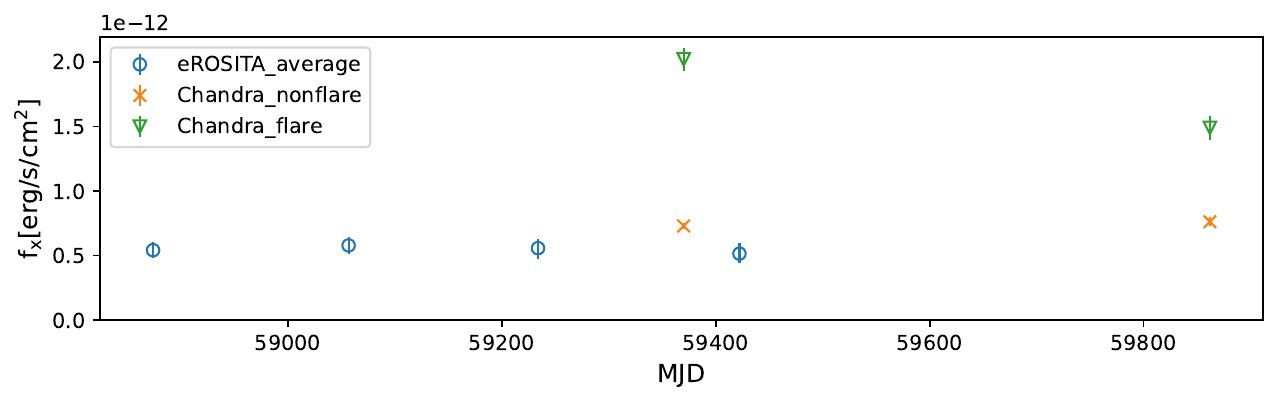}
        \caption{\sredit{Long-term X-ray} Variability of the \sredit{combined} BC components. The average flux taken from eROSITA eRASS1:4 (blue) every six months during the years  2020-2021. The Chandra observation from year 2021 was reported in \cite{Brown2022} and year 2022 from this work during the flare (green) and non-flare (orange) time.}
    \label{fig:monitoring_bc}
\end{figure*}

In this section, we present the results derived from the analysis of X-ray \sredit{observations} obtained from Chandra and eROSITA. These observations provided X-ray properties, and the activity levels of the sources, as detailed in Section \ref{sec:xrayproperties}. Additionally, the eROSITA observation revealed X-ray variability spanning from eRASS1 to eRASS4, occurring two years before the Chandra observations. This variability is discussed in Section \ref{sec:xrayvariability}. We note that eROSITA is unable to distinguish the individual components in this system. \sredit{We place the \sredit{X-ray-to-bolometric ratio} in context of the Rossby number in Section \ref{subsec:Activity_levels}}. Lastly, the results from \sredit{atmospheric} model predictions are presented in Section \ref{subsec:Model_prediction}.

\subsection{X-ray properties from spectral analysis} \label{sec:xrayproperties}

The spectral fits and parameter constraints from \texttt{BXA} and the plasma model \texttt{APEC} are presented in Fig. \ref{fig:LTT1445ABC_spec}. For the spectral analysis of the A component, we combined all observations for a total duration of $50$\,ks, resulting in the top panel of the figure.

Regarding the BC components, we specifically selected data from flaring periods, covering the time intervals of 1800-4000 s in the second observation and 14000-18000 s in the third observation in Fig.\ref{fig:chandra_lightcurve}. The corresponding spectrum from these flare periods is displayed in the middle panel of Figure~\ref{fig:LTT1445ABC_spec}. The remaining period from all three observations is categorized as quasi-quiescent, and the respective spectrum is depicted in the bottom panel of Figure~\ref{fig:LTT1445ABC_spec}. Our spectral model incorporated the temperature distribution for the flare and quasi-quiescent periods, respectively. Our \texttt{APEC} model \sredit{models} a temperature distribution (kTdist) instead of a single-point temperature. An advantage of using a temperature distribution is that it captures the behavior of the plasma temperature better than a single point, resulting in the distribution shape of the Differential Emission Measure (DEM) shown in the left panel of Fig.\ref{fig:LTT1445ABC_spec}.

In addition to the fitted spectral profile, we display the temperature distribution of each dataset \sredit{obtained with the 10-component grid model}. The quiet period of A and the quasi-quiescent period of BC show \sredit{low emission at} high temperatures, while the distribution of the BC flare \sredit{extends to higher temperatures, as shown in the right panels of Fig. \ref{fig:LTT1445ABC_spec}.}

Despite the low counts, we allow the metal abundance as a free parameter in the \texttt{APEC} model. The obtained value for A's abundances is 0.66$\pm$0.23 relative to the solar abundances. For BC during quasi-quiescent, it is 0.141$\pm$0.045, and for BC during a flare, it is 0.49$\pm$0.21.

A detailed examination of the X-ray characteristics is provided in Table \ref{tab:LTT1445A_properties} for the A component and Table \ref{tab:ltt1445bc_properties} for the BC component. The spectral analysis results demonstrate that each component exhibited similar X-ray luminosity levels both during A's observation and throughout the BC observations. However, the BC component displayed X-ray luminosity levels approximately one order of magnitude higher than those observed for the quiescent A.

\subsection{X-ray Variability} \label{sec:xrayvariability}

The BC components exhibit variability from both B and C at a similar level, as indicated in Table~\ref{tab:ltt1445bc_properties}. Flare periods are observed for both B and C, as illustrated in the light curve in Fig.~\ref{fig:chandra_lightcurve}. \cite{Foster2022} reported a significant signal originating from the BC component from eRASS1.

However, due to the instrument's spatial resolution, eROSITA cannot distinguish between the three components, and we report the eROSITA fluxes as totals for the entire system and assume the total is dominated by BC. Fig.~\ref{fig:monitoring_bc} presents the flux from eRASS1 to eRASS4, along with the observed Chandra flux during flare and non-flare times for BC. We converted the 0.6-2.3 keV eROSITA count rates to fluxes with an APEC model, assuming a plasma temperature (kT) of 0.5 keV absorbed by $\log N_\mathrm{H}/\mathrm{cm}^2=19.5$ following \cite{Brown2022}. If we adopt 1.0 keV instead, the flux becomes 6.6\% lower. For Chandra, we obtain fluxes in the 0.6-2.3 keV range from the spectral fit shown in Fig.~\ref{fig:LTT1445ABC_spec}. A caveat here is that as of 2020, Chandra is most sensitive from 0.9-7 keV, while eROSITA has most sensitive energy range from 0.3-2.3 keV. 
This difference could lead to systematic discrepancies\sredit{, see Figure 9 in \cite{Predehl2021} for details}. This could explain the offset between the lowest Chandra flux measurement to the eROSITA measurements.

\subsection{X-ray Activity Levels} \label{subsec:Activity_levels}

\begin{figure}[h!]
    \includegraphics[width=\columnwidth]{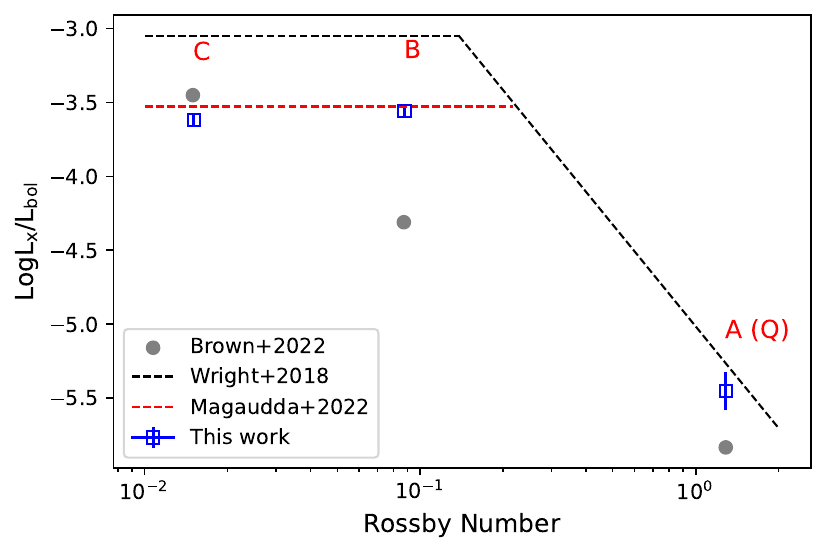}
    \caption{The updated \sredit{X-ray-to-bolometric ratio} \sredit{and Rossby numbers. Our results are in blue and those from \cite{Brown2022} in gray. They are compared} with the \cite{Wright2018} model. \sredit{\cite{Wright2018} saturation level appears excessively high for low-mass M stars, whereas \cite{Magaudda2022} reports a value close to -3.5 (their Fig.9).}}
    \label{fig:rossby}
\end{figure}

We \sredit{compare} the observed X-ray activity with the standard behavior exhibited by M dwarfs in Fig.~\ref{fig:rossby}. The Rossby number ($R_0=P_{\text{rot}}/\tau$), representing the ratio of the rotation period ($P_{\text{rot}}$) to the convection turnover time ($\tau$), is calculated following \cite{Wright_2011}, as expressed in equation~5 from \cite{Wright2018}. These investigations, primarily focused on F-G-K-M dwarfs, utilized ROSAT data for their analyses.

\sredit{The} rotation periods for A, B, and C, are adopted from \cite{Winters2022} as 85, 6.7, and 1.4 days, respectively. From light curve periodograms, \cite{Winters2022} were able to \sredit{identify} a 1.4-day peak consistent with the rotational broadening observed in the spectrum of C. 

The smaller rotational broadening of B was associated with a  \sredit{marginal} 6.7-day periodicity peak. From a broad and weak periodogram peak from 50 to 100 days they suggest a rotation period of $85\pm20$ days for A. The radial velocity campaign of \cite{Lavie2022} for A finds a significant periodicity of 72 days in line widths, which they interpret as the rotation period for A.  \sredit{These authors note} that the long rotation periods of A and the short periods of BC are at odds with the stars being born together assuming common rotation spin-down models. We adopt 85 days for consistency with the previous study of \cite{Brown2022}. 

\sredit{Since we adopt the same periods as \cite{Brown2022}, our Rossby numbers are the same, namely for 
components A, B, and C $R_o$ is} 1.3$\pm$1.2, 0.09$\pm$0.08, and 0.015$\pm$0.015, respectively.

We have updated the X-ray bolometric luminosity ratios for the LTT1445 X-ray system. This involves calculating the luminosity ratio using the bolometric luminosity data from \cite{Winters2019} along with our observed \sredit{time-averaged}  X-ray luminosities. Notably, the inclusion of the observed flare in B enhances the alignment of the data points with the anticipated relation outlined by \cite{Wright2018}. 

During the Chandra observations in late 2022 (or Q4 2022), the A component was quiescent. \sredit{Initially C was active and B quasi-quiescent, while in the second half of the light curves shown in Fig.~\ref{fig:chandra_lightcurve}, B is active and C appears quiescent}. The updated activity relation is shown in Fig.\ref{fig:rossby}.

\subsection{Model prediction} \label{subsec:Model_prediction}
\subsubsection{Stellar Evolution}

\begin{figure}[h!]
    \centering
    \includegraphics[width=\columnwidth]{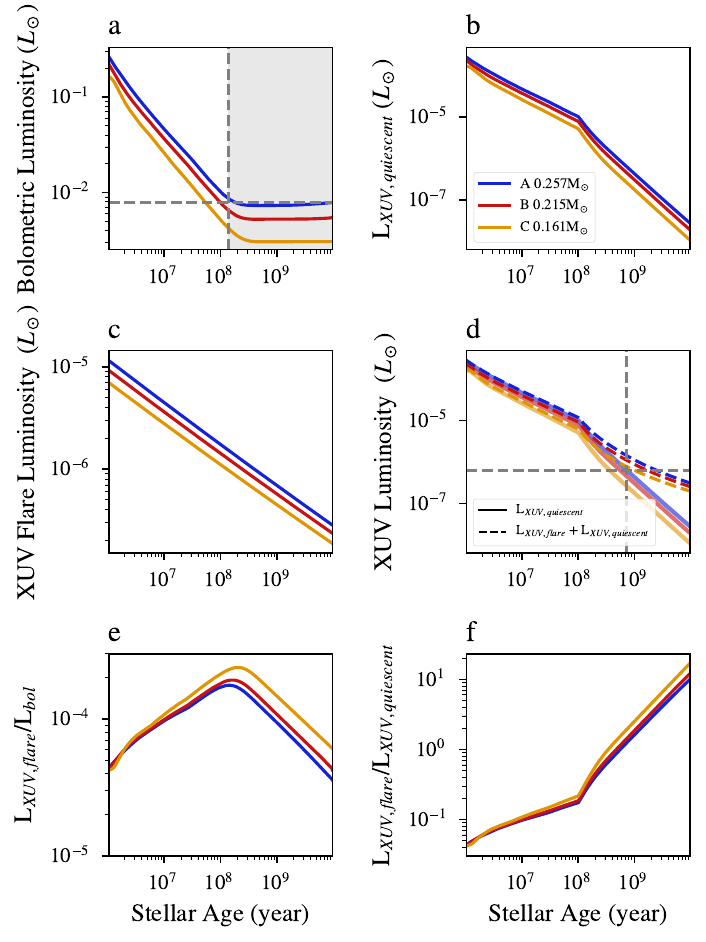}
    \caption{The stellar evolution track based on \cite{Baraffe2015} model of the three stars in the system. Panels (a) and (b) display the bolometric luminosity and the XUV quiescent luminosity. The observed bolometric luminosity and XUV luminosity were used to determine the A component. Panels (c) and (d) display the XUV luminosity from flares and the total XUV luminosity overlaid by the gray dashed line indicating the average observed X-ray flux + estimated EUV flux in (d). The ratio of XUV luminosity from flares to bolometric luminosity is displayed in panel (e), and the ratio of XUV luminosity from flares to XUV quiescent luminosity is displayed in panel (f). }
    \label{fig:StellarEvol}
\end{figure}

\sredit{We use simulations to place the X-ray observations of these M dwarfs in context. For age estimation,} we provide an overview of stellar evolution using the \texttt{MDwarfLuminosity} \citep{Amaral2022} module of the \texttt{VPLanet} package. This is illustrated in Fig. \ref{fig:StellarEvol}, where we explore the luminosity and structural changes that stars undergo from their formation until the culmination of the main sequence phase. Furthermore, this model includes X-ray flares, quiescent fluxes, and their impact on the planets\sredit{ as described already in section \ref{subsec:atmo_model}. In the evolutionary model, the measured bolometric luminosity \citep{Winters2019} is placed with gray dashed line in panel a of Fig. \ref{fig:StellarEvol}, suggesting a stellar age of > 0.14 Gyr once the star entered the main sequence phase. Similarly, the measured X-ray flux combined with the estimated EUV flux in panel d suggests the stellar age of $\sim$ 1 Gyr. } 

\subsubsection{Atmospheric Escape}
\sredit{In this section, we present the predicted impact of XUV-driven atmospheric escape. As described in \ref{subsec:atmo_model}, for} terrestrial planets, \texttt{VPLanet} assumes atmospheres dominated by water vapor rather than H/He. This module follows a runaway greenhouse scenario, where atmospheric escape occurs only when the total incident flux on the planet surpasses the runaway greenhouse threshold \citep{Kopparapu2013}. In the context of water and oxygen loss, this occurs after the removal of the Hydrogen envelope. Subsequently, XUV photons start the dissociation of water molecules, leading to increased Hydrogen escape and, in certain scenarios, oxygen escape.

\begin{figure*}[h!]
    \centering
    \includegraphics[width=\textwidth]{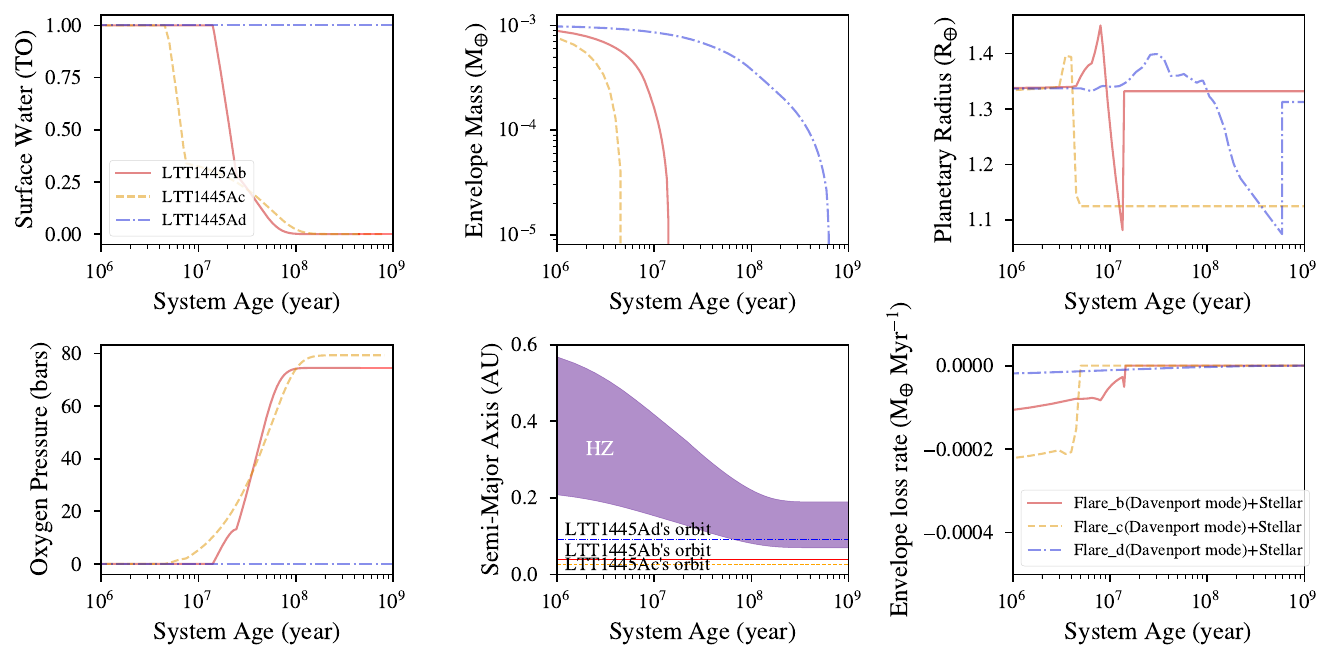}
    \caption{Atmospheric escape over time of the LTT 1445A. The panels show: (top left) the surface water starting from water abundance of 1.0 TO, (top center) Hydrogen envelop mass loss of the three planets over time, (top right) the evolution of planetary radius, (bottom left) the amount of oxygen buildup in the atmosphere due to the XUV radiation, (bottom central) position of each planet from the habitable zone (HZ), (bottom right) envelope loss rate.}
        \label{fig:atmo_escape}
\end{figure*}

Figure \ref{fig:atmo_escape} illustrates the outcomes of our simulations, indicating that planet c's Hydrogen envelope was stripped away after 4.7 million years, planet b's envelope endured for 14.1 million years, and planet d's envelope after 623.8 from the initial age, \sredit{set at 5 Myr when the disk disperses. This is a pessimistic scenario, as the early high XUV radiation (see Fig.\ref{fig:StellarEvol}) strips both water and oxygen from the planet. Running a more optimistic scenario with a start time at e.g. 200 Myr retains slightly more water. However, the results are within the same order of magnitude, because while water is dissociated earlier, the oxygen atmospheric evaporation is only partial.}

Notably, planet d, initially thought to be within the habitable zone (HZ) according to \cite{Lavie2022}, \sredit{is in the 'Recent Venus' zone, which is in between 'Recent Venus' and 'Runaway Greenhouse' HZ limit (Fig.\ref{fig:habitable_zone}) when the stellar age is considered. Any planet located beyond the dotted yellow line, which represents the Recent Venus HZ limit, is empirically derived from the understanding that Venus has no liquid water on its surface for at least the last one billion years \citep{Solomon1991}. }This HZ planet suggests the potential retention of surface water, while for the other planets, abiotic oxygen \sredit{accumulates while the water reservoir dissipates}. However, \cite{Amaral2022} suggested that water content is more significantly influenced by stellar mass than \sredit{by} planetary mass when water loss ceases upon reaching the HZ. 

\section{Discussion} \label{sec:discussion}

The understanding of the intricate interplay between stars and their planets is pivotal for advancing future observations of rocky planets and refining the characterization of their atmospheres. In this section, various aspects of the star-planet interaction in the triple-star system LTT 1445 are explored, incorporating observational parameters to enhance and constrain theoretical models. This includes considerations of stellar evolution, the estimation of planetary mass loss, and atmospheric escape.

From the simulated evolution of the exoplanet atmospheres, the predicted current state can be read off at the time corresponding to the current age of the star. However, determining the age of M dwarfs poses challenges \citep{Stelzer2016, Magaudda2020}, and the limited observation time provides only a narrow snapshot of the system. Despite these challenges, we attempt to determine the age through three methods: 1) evolutionary modeling \citep{Baraffe2015, Davenport2019, Amaral2022} through \texttt{VPLanet} package, 2) X-ray activity-age relation \citep{Magaudda2020}, and 3) optical-near infrared activity-age relation \citep{Engle2023}.

\sredit{First, we integrated observed bolometric \sredit{luminosities from} \cite{Winters2019} with the \cite{Baraffe2015} model. It is noteworthy that this model grid lacks XUV evolution, which is not well-constrained for M-dwarf stars. The XUV flux of \texttt{STELLAR} package follows \cite{Bolmont2017_waterloss}. \sredit{The evolution of the bolometric luminosity halts} after reaching the main sequence at 0.14 Gyrs at a luminosity comparable to that of the \sredit{exoplanet host star LTT1445A}. Therefore, the star is on the main-sequence and older than 0.14 Gyr (gray area in panel a of Fig.\ref{fig:StellarEvol}). For the XUV luminosity panels c and d, in the panel c of Fig.\ref{fig:StellarEvol} if we consider only the observed X-ray flare luminosity of A from \cite{Brown2022}, without the UV component overlaid on the model, the result is approximately 
\sredit{2.0\,Gyr}. In panel d, we sum the observed X-ray luminosity average during quiescence ($\mathrm{logL_X\sim 26}$ erg/s) with the EUV flux ($\mathrm{logL_{EUV}\sim27}$ erg/s) estimation following \citep{SanzForcada2011}, which is similar to the GALEX flux from \cite{Stelzer2013}. Comparing the total XUV luminosity (dashed horizontal line in panel d of Fig.~\ref{fig:StellarEvol}) to the blue model curve yields an age of 0.73 Gyr. }

\sredit{Second, the X-ray luminosity-rotation period relation (Figure 7 of \citet{Magaudda2020}, Figure 8 of \citet{Magaudda2022}) for an X-ray luminosity level of $\mathrm{logL_X=26}$ erg/s suggests a rotation period on the order of 100 days. This is consistent with rotation constraints of \cite{Winters2022} and \cite{Lavie2022}. 
Both observational relations of rotation period and age (from binaries with white dwarfs companions) by \cite{Engle2023} and \cite{Magaudda2020} suggest an age of \sredit{$\mathrm{> 2.5\,Gyr}$}. 
}

\sredit{Lastly, plugging the rotation period of 85 days into the period-age relation derived from a TESS M-dwarf sample (equation 3 in \cite{Engle2023}), estimates an age of 2.37 Gyr for the A component. To summarize, the estimates above suggest that the age of the A component is in the range of 1 to 2.5 Gyr. For the simulations considered below, the exact age is not critical, as the atmosphere evolution stabilizes after approximately 100 Myr. }

\sredit{The differences in rotation periods for A, B and C are unexpectedly large when compared to isolated stars of the same age. This was already noted in \cite{Lavie2022}. The age estimates for BC provided in \cite{Brown2022} and the Hubble Space Telescope (HST) observation detailed in \cite{Pass2023} suggest that both BC components may represent a younger, faster-rotating star compared to the Sun. }
\sredit{This raises the question how the three stellar components can have such different ages. Differences in stellar mass and stellar type can be understood from theoretical considerations} encompassing stellar evolution, interactions, three-body dynamics, and the initial population in stellar triples, which collectively contribute to variations in mass and stellar types. Simulations by \cite{Toonen2020} indicate that stellar interactions are common in triples. In contrast to binary populations, the proportion of systems capable of undergoing mass transfer is approximately 2-3 times greater in triple systems, \sredit{potentially} resulting in varied mass within the system. Insights into planet formation in complex stellar environments can be \sredit{gained} from studying \sredit{protoplanetary disks in young star systems}. \cite{Ronco2021} investigated a nearly 10 Myr old disk in the HD 98800 multiple system, which still \sredit{retains} significant amounts of gas. The study compared the evolution of gas disks in hierarchical triple-star systems and circumbinary systems. It revealed that in triple-star systems, gas surface density profiles evolve slowly, leading to higher midplane temperatures and aspect ratios compared to circumbinary systems. This could potentially facilitate planet formation in triple-star systems. Recent discoveries of protoplanetary disks in multiple-star systems also suggest that some of these disks can remain intact for much longer periods; for instance, \cite{Zagaria2021}, observed that the motion of dust within circumstellar disks around one of the stars in a binary star system exhibits a higher drift rate compared to disks unaffected by an outer stellar companion.

\sredit{For a worst-case scenario of atmosphere loss, we utilize the high} X-ray flux of LTT1445A during the observed flare period in \cite{Brown2022}. The determined mass loss rates are approximately $\mathrm{1.41 \times 10^8 g/s}$ for planet b, $\mathrm{3.22 \times 10^9 g/s}$ for planet c, and $\mathrm{2.93 \times 10^9 g/s}$ for planet d, considering a $R_{pl}$ value of 1.29 $R_\earth$ for planet d.


\sredit{Within the planetary ensemble that surrounds the triplet star system \citep{Cuntz2022}, various orbital configurations may exist. These configurations range from planets orbiting a single star to those in binary \citep{Eggl2018,Ciardi2021} or circumtriple orbits \citep{smallwood2021} within the system. Although the LTT1445A's planets are the simplest configuration, we explore the potential exposure of these planets to flares originating not only from their host star but also from neighboring stars.}

We assess the impact of BC on the planets orbiting A. The distance from A to BC is 34 AU, whereas the orbital distance of the planets is less than 0.1 AU. Consequently, due to the significant distance, the X-rays originating from A are approximately 400$^2$ times more influential. Despite the largest observed flare from BC (luminosity of $10^{28.15}\,\mathrm{erg/s}$, Table~\ref{tab:ltt1445bc_properties}), its impact on the planets is negligible compared to the irradiation from the most quiescent X-ray levels of A (luminosity of $10^{25.5}\,\mathrm{erg/s}$, Table~\ref{tab:LTT1445A_properties}).
%
%
%

\sredit{Fig. \ref{fig:atmo_escape} presents the outcomes from \texttt{VPLanet}, including surface water loss, oxygen buildup, Hydrogen envelope mass loss, and the evolution of planetary radius over time. The results suggest that planet LTT 1445Ad could retain a water surface\sredit{. Planets Ab and Ac may accumulate} abiotic oxygen up to pressures of 70-80 bars if the system initially possessed 1.0 Terrestrial Ocean (TO), as depicted in the left column of Fig. \ref{fig:atmo_escape}. Our simulation assumes that the evolution began around 5 Myr. 

In the top left corner of Fig. \ref{fig:atmo_escape}, we observe the rapid loss of surface water of planets c and b, which are closer to the star before 11 Myr. Subsequently, abiotic oxygen buildup commenced and remained constant over 1 Gyr. However, a more accurate evaluation of this assumption requires precise measurements of water abundance to refine the model.}

\section{Summary} \label{sec:summary}

This study presents a detailed analysis of the long-term high-energy environment of LTT1445. The dataset encompasses three recent Chandra observations, coupled with continuous monitoring from eROSITA eRASS1 to eRASS4. The Chandra observations effectively differentiate coronal X-ray emissions from all three stars in the LTT 1445 system, including the exo\-planet host star, LTT 1445A. Our primary findings are as follows:

\begin{itemize}
  \item Concerning the star's X-ray activity, LTT 1445C emerges as \sredit{the dominant X-ray source}, with \sredit{secondary} contributions from LTT 1445B. The study confirms that LTT 1445A, characterized by slow rotation, exhibits no significant flare activity in our observational dataset.
  \item The planets orbiting A receive irradiation from all M dwarfs. However, our findings suggest that the X-ray emission from the LTT 1445BC components does not pose a greater threat to the planets orbiting LTT 1445A than the emissions from A itself.
  \item If the system's evolution begins with a water abundance of at least 1.0 TO, LTT 1445Ad could have an atmosphere. \sredit{Moreover, planet d appears capable of retaining water over approximately a 1 Gyr time scale, making it suitable for future observations.}
  \item X-ray irradiation causes non-biotic oxygen build-up in the atmosphere of the planets in this nearby system.
  \sredit{\item This makes LTT1445Ad, aided by the proximity and apparent brightness of LTT1445A, the most promising target for future atmosphere studies searching for water and molecular oxygen. If oxygen is detected on planet d, the simulation performed here predicts that it is not of abiotic origin. In any case, future observation should first constrain the water content in the system.}
\end{itemize}

\sredit{Our simulations determine the current presence of water and oxygen in the planetary atmospheres. The simulations rely on assumptions that are well-determined, such as the stellar mass and planetary mass, and others that are less clear, such as an initial water abundance at the start of the simulations. It is important to note that processes such as radiative cooling from substances such as $\mathrm{CO_2}$, tidal forces, planetary magnetic fields, or coronal mass ejections (CMEs) were not considered. Whether these processes make a minor contribution or significantly alter the overall picture will be studied in the future. The model should be tested and refined, including with accurate measurements of water abundance in the planetary atmosphere e.g. with James Webb Space Telescope (JWST).}

\sredit{For atmospheres harboring $\mathrm{O_2}$ from transiting exoplanets, LTT1445A's planets are prime candidates. In our model, under the assumption of a\sredit{n initial} water reservoir, $\mathrm{O_2}$ is abundantly produced by photo-dissociation. While the $\mathrm{O_2}$ is abiotic, the freed $\mathrm{O_2}$ has the potential to activate further reactions. The presence of oxygen could be tested using novel exoplanet transmission spectroscopy instrumentation on Extremely Large Telescopes.} For example, FIOS \citep{Ben_Ami_2018}, where an ultra-high-resolution Fabry-Perot-based instrument boosts spectral resolution, separating the $\mathrm{O_2}$ absorption spectrum's peaks and troughs, which reduces the observing time by 35\% \citep{Currie2023, Hardegree2023}. \sredit{A FIOS prototype was demonstrated recently with on-sky observations with telluric oxygen by \cite{Rukdee2023}}. For direct imaging of strongly illuminated planets, the Ultra-fast AO techNology Determination for Exoplanet imageRs from the GROUND (UNDERGROUND) \citep{Fowler2023} is planning to observe the reflected light using a combination of a coronagraph and Virtually Imaged Phase Arrays (VIPA) high-resolution spectrograph. These observations may validate the predictions presented here and potentially contribute to the discovery of an inhabited exoplanet. To further refine our simulation assumptions, it is recommended that future panchromatic observations be conducted to fully understand the environmental context, utilizing space missions such as the James Webb Space Telescope for identifying the planetary water content and Athena for the assessment of the high-energy environment. 


\section*{Software}
This work makes use of the following package: \texttt{matplotlib} \citep{Hunter2007}, \texttt{scipy} \citep{2020SciPy_NMeth}, \texttt{astropy} \citep{astropy:2013,astropy:2018,astropy:2022}, \texttt{BXA} \citep{Buchner2014}, \texttt{UltraNest} \citep{Buchner2021}, \texttt{CIAO/Sherpa} \citep{Fruscione2006}, \texttt{eSASS} \citep{Brunner2022} and \texttt{VPLanet} \citep{Barnes2020} including package \texttt{FLARE} \citep{Amaral2022}.

\begin{acknowledgements}
This work is based on data from eROSITA, the soft X-ray instrument aboard SRG, a joint Russian-German science mission supported by the Russian Space Agency (Roskosmos), in the interests of the Russian Academy of Sciences represented by its Space Research Institute (IKI), and the Deutsches Zentrum für Luft- und Raumfahrt (DLR). The SRG spacecraft was built by Lavochkin Association (NPOL) and its subcontractors, and is operated by NPOL with support from the Max Planck Institute for Extraterrestrial Physics (MPE). 

The development and construction of the eROSITA X-ray instrument was led by MPE, with contributions from the Dr. Karl Remeis Observatory Bamberg \& ECAP (FAU Erlangen-Nuernberg), the University of Hamburg Observatory, the Leibniz Institute for Astrophysics Potsdam (AIP), and the Institute for Astronomy and Astrophysics of the University of Tübingen, with the support of DLR and the Max Planck Society. The Argelander Institute for Astronomy of the University of Bonn and the Ludwig Maximilians Universität Munich also participated in the science preparation for eROSITA. 

The eROSITA data shown here were processed using the eSASS/NRTA software system developed by the German eROSITA consortium.

SR thanks Rory Barnes, Laura do Amaral and Ludmila Carone for the \texttt{VPLanet} package tutorial during the VPLanet workshop. \sredit{SR also thanks} Enza Magaudda for constructive conversations regarding the age estimation.
\end{acknowledgements}

%
%

\bibliography{ref}
\bibliographystyle{aa}

\end{document}